\documentclass[twocolumn, aps,graphicx,prb, showpacs]{revtex4}
\usepackage{lipsum}
\usepackage{graphicx}
\usepackage{anyfontsize}

\usepackage[a4paper, total={6in, 8in},margin=0.5in]{geometry}
\usepackage[font=scriptsize,justification=justified]{caption}
\usepackage{subfigure}
\usepackage[export]{adjustbox}
\usepackage{float}
\usepackage{placeins}

\begin{document}
\title{A single-atom 3D sub-attonewton force sensor}

\author{V. Bl\={u}ms$^1$, M. Piotrowski$^{1,2}$, M. I. Hussain$^1$, B. G. Norton$^1$, S. C. Connell$^{1,2}$, S. Gensemer$^{1,2}$,M. Lobino$^{1,3}$, E. W. Streed$^{1,4\star}$ \\
$^1$\textit{Centre for Quantum Dynamics, Griffith University, Brisbane, 4111 QLD, Australia.}\\
$^2$\textit{CSIRO Manufacturing, Pullenvale, 4069 QLD, Australia}\\
$^3$\textit{Queensland Micro and Nanotechnology Centre, Griffith University, Brisbane, QLD 4111, Australia}\\
$^4$\textit{Institute for Glycomics, Griffith University, Gold Coast, QLD 4222, Australia}\\
\textit{$^\star$email: e.streed@griffith.edu.au}}


\begin{abstract}
All physical interactions are mediated by forces. Ultra-sensitive force measurements are therefore a crucial tool for investigating the fundamental physics of magnetic \cite{Rugar-Chui-2004,Bleszynski-Jayich-Harris-2009}, atomic\cite{Gangloff-2015}, quantum \cite{Mohideen-Roy-1998,Chan-Capasso-2001}, and surface \cite{deLepinay-AFM-2D-Nanowire-Forces,Rossi-AFM-2D-Nanowire-Forces-other} phenomena. Laser cooled trapped atomic ions are a well controlled quantum system and a standard platform for precision metrology\cite{Chou_Al_clock}. Their low mass, strong Coulomb interaction, and readily detectable fluorescence signal make trapped ions favourable for performing high-sensitivity force measurements\cite{Biercuk-Bollinger-2010,Shaniv:2017}. Here we demonstrate a three-dimensional sub-attonewton sensitivity force sensor based on super-resolution imaging of the fluorescence from a single laser cooled $^{174}$Yb$^+$ ion in a Paul trap. The force is detected by measuring the net ion displacement with nanometer precision\cite{Wong-Monroe-2015}, and does not rely on mechanical oscillation. Observed sensitivities were 372$\pm$9$_{\mbox{stat}}$,  347$\pm$12$_{\mbox{sys}}\pm$14$_{\mbox{stat}}$, and 808$\pm$29$_{\mbox{sys}}\pm$42$_{\mbox{stat}}$ ~zN/$\sqrt{\mathrm{Hz}}$ in the three dimensions, corresponding to 24x, 87x, and 21x of the quantum limit. We independently verified the accuracy of this apparatus by measuring a light pressure force of 95~zN on the ion, an important systematic effect in any optically based force sensor. This technique can be applied for sensing DC or low frequency forces external to the trap\cite{Harlander-ion-trap-dielectric-surface-charging} or internally from a co-trapped biomolecule\cite{Trypogeorgos-Cotrapping} or nanoparticle\cite{Kuhlicke-nanodiamond-ion-trap}.
\end{abstract}
\date{\today}

\maketitle

The development of high-resolution imaging of laser cooled trapped ions\cite{Jechow-11, Maiwald-12, Wong-Monroe-2015} opened the possibility for the realisation of an ion based sensor that could resolve an external force in all three directions using a single atomic ion. In a harmonic potential Hooke's law $F_i= k_i \Delta x_i$ allows us to convert displacement measurements $\Delta x_i$ into a force measurements $F_i$ through the associated spring constants $k_i$. Forces parallel to the image plane are detected by measuring the displacement of the ion's centroid while forces applied orthogonal to that plane in the focusing direction, parallel to the optical axis of the imaging apparatus, are measured from a change in the width of a slightly defocused ion image. While the resolution of a fluorescence image is limited by the wavelength of light, the exact centroid location and width can be determined to a much greater precision through super-resolution imaging techniques\cite{Thompson-FIONA,Pertsinidis-sub-nm-super-resolution}. Doppler velocimetry based force sensing in Penning\cite{Biercuk-Bollinger-2010} and Paul\cite{Shaniv:2017} traps has demonstrated sensitivities down to 28 yN/$\sqrt{\mathrm{Hz}}$, however this is limited to one dimension and a narrow frequency band around a driven oscillation. Force measurement through imaging is a broadband rather than a narrowband sensing technique, and without a fundamental lower frequency limit to detection.

The quantum limits to force sensing through imaging are necessarily linked to both the imaging resolution and the detected photon throughput. For a displacement accuracy of $\delta x_i$ we can measure a force with uncertainty $\delta F_i= k_i \delta x_i$, where $k_i$ are the spring constants. In ion traps the long-term stability of the spring constants can be below $<10^{-5}$ level with an optimised hardware design\cite{Johnson-Ion-Trap-Freq-Active-Stab}, making their uncertainty contribution negligible. Assuming a Gaussian mode for our imaging\cite{Thompson-FIONA}, and ignoring technical limitations from pixelisation and background noise, for $N$ detected photons at wavelength $\lambda$ with a numerical aperture of $\mbox{NA}$ the transverse uncertainty in the image plane is at best $\delta x_i =  \lambda/\left(\pi\mbox{NA} \sqrt{N}\right)$ and $\delta z= 2 \lambda/ \left(\pi \mbox{NA}^2 \sqrt{N}\right)$ in the focus direction (see Methods). In real experiments the effective quantum limits are reduced due to imaging imperfections.

\begin{figure}[h!]

\includegraphics[scale=1]{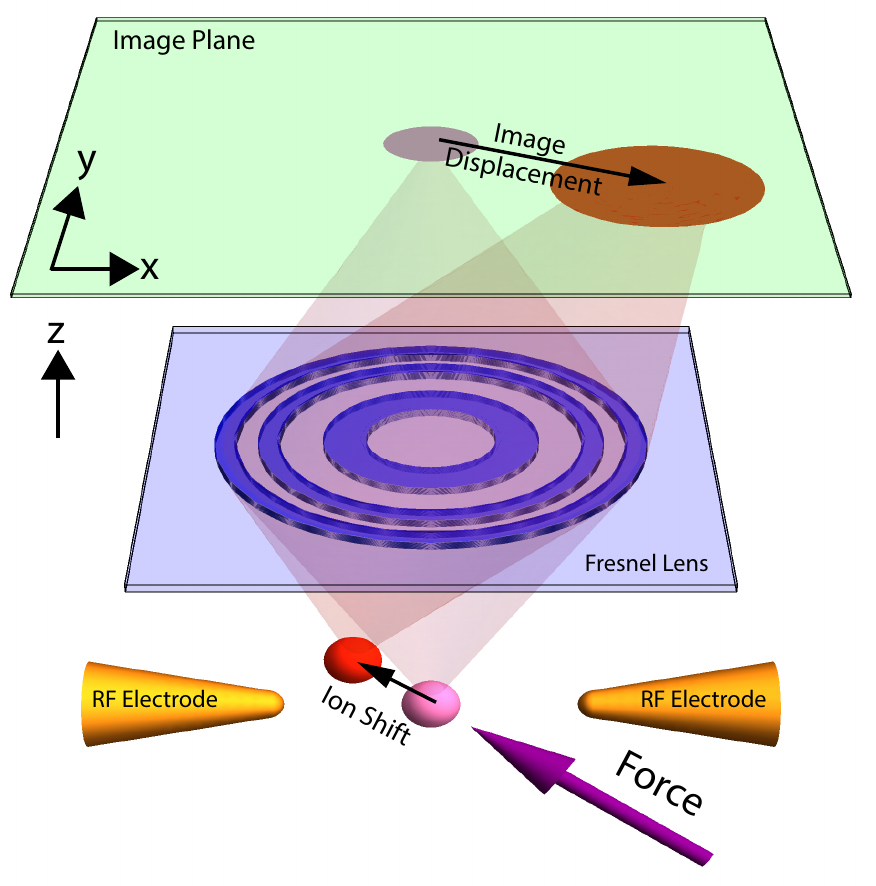}
\captionsetup[subfigure]{justification=justified,singlelinecheck=false}
\caption{\textbf{Experimental configuration.} An externally applied force (purple arrow) displaces a trapped $^{174}$Yb$^+$ ion. The image formed by the phase Fresnel lens is shifted in two dimensions. Displacement along the focal axis alters the image spot size. For clarity illustrated trap dimensions are not to scale. Radiofrequency (RF) electrode needle spacing 300$\mu$m, lens focal length 3 mm.}
\label{fig_1}
\end{figure}

\begin{figure}[t]
\begin{center}
\includegraphics[scale=1]{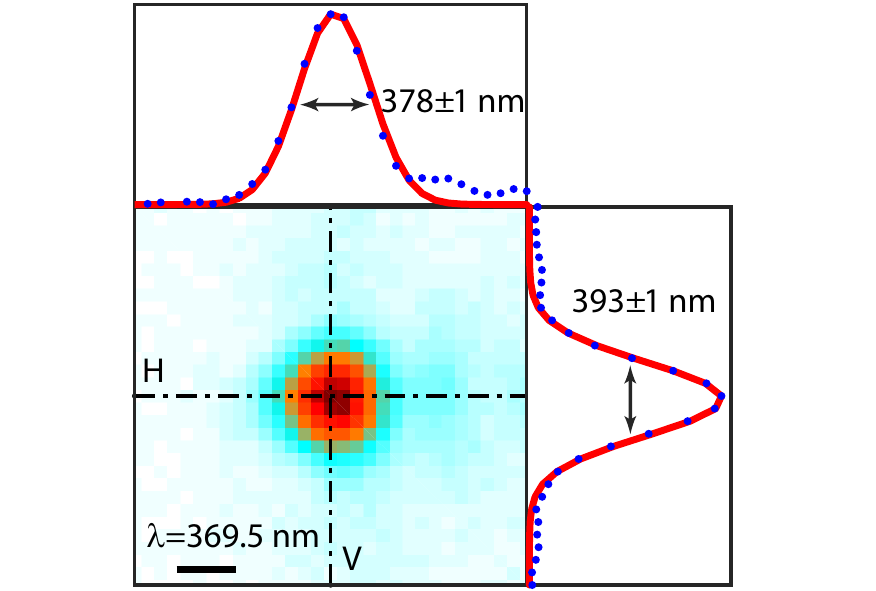}
\caption{\textbf{Ion image.} Wavelength-scale ion image with Gaussian fitting and full width at half maximum sizes (FWHM) for the horizontal and vertical slices.}
\label{fig_2}
\end{center}
\end{figure}

Our ion trapping apparatus\cite{Streed-11,Jechow-11,Streed-Kielpinski-2012-Single-atom-absorption-imaging} is shown in Fig.~\ref{fig_1} and images the fluorescence from a single laser cooled $^{174}$Yb$^+$ (see Methods) at $\lambda=369.5$ nm. A typical in-focus image of our trapped ion is shown in Fig.~\ref{fig_2} with full-width half maximum (FWHM) diameters of 378$\pm$1~nm and 393$\pm$1~nm in $x$ and $y$. With a typical 20~s exposure time and accounting for systematic drifts (see Methods), we are able to determine the ion's centroid position with a precision of 2.8~nm and 10~nm in the $x$ and $y$ directions respectively. By defocusing the camera and calibrated measurement of changes to the spot width (see Methods) we were able to determine shifts along the $z$ direction to within 24~nm.

\begin{figure}[b!]
\centering
\includegraphics[scale=1]{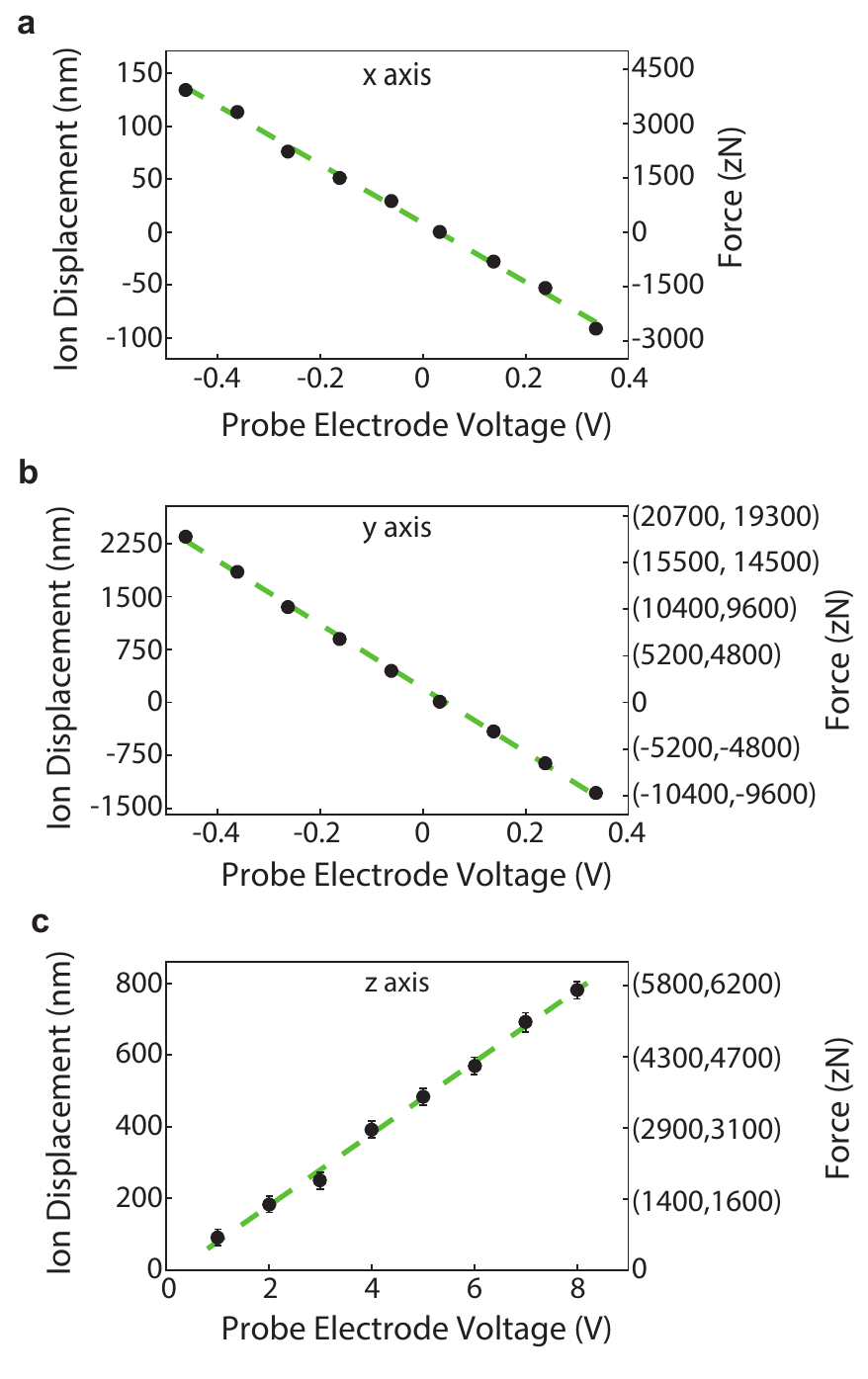}
\caption{\textbf{Electrostatic force detection and ion movement in three axes.} x-axis and y-axis refer to camera plane and z-axis refers to the optical axis of the imaging system. \textbf{a}, Ion displacement and applied force as a function of the external voltage in the x-axis. This axis correspond to the orientation of the trapping needles and has the strongest confinement. \textbf{b}, Same as \textbf{a} but for the y-axis. The use of ranges in the Force vertical axis of the plot represent the constrained systematic uncertainty due to unknown orientation of the trap principal axis with respect to our optical system. Error bars in \textbf{a} and \textbf{b} are smaller than the dots. \textbf{c}, Same as \textbf{b} but for the z-axis.}
\label{fig:shift_3_axes}
\end{figure}

To characterise the system's ability to measure small forces, we performed a series of measurements of ion displacements under the influence of different external electrostatic fields (see Methods). Figure~\ref{fig:shift_3_axes}a and b shows the measured forces on the ion as a function of the applied voltage together with the associated displacement in the $x$ and $y$ directions. Considering the position resolution of our system and the 20~s acquisition time, we measured a sensitivity S$_x$~=~372$\pm$9~zN$/\sqrt{\mathrm{Hz}}$ and S$_y$~=~(335, 359)$\pm$14~zN$/\sqrt{\mathrm{Hz}}$, where the range in the value of S$_y$ is due to systematic error associated with the unknown trap axis orientation. The large difference in displacement between Fig.~\ref{fig:shift_3_axes}a and b is consistent with the difference in spring constant between the stronger confinement in k$_x$ (along the trapping needle axis) and the weaker confinement in k$_y$ (perpendicular to the trapping needle axis) combined with the position of the electrode generating the electrostatic field. Force measurement in the $z$ direction is shown in Fig.~\ref{fig:shift_3_axes}c together with the corresponding change in position as a function of the external voltage from which we calculated a force sensitivity S$_z$~=~(779, 836)$\pm$42~zN$/\sqrt{\mathrm{Hz}}$. Figure~\ref{fig:shift_3_axes} shows the linearity of the displacement as a function of the applied voltage in all three directions as expected for an harmonic potential for forces up to 20~aN.

To independently verify the accuracy of our force sensing technique we measured the scattering light force from our cooling laser on the ion. Absorption generates a force in the direction of the laser beam $\mathrm{\vec{F}=\gamma\hbar \vec{k}}$, proportional to the scattering rate $\mathrm{\gamma}$ and the photon momentum $\mathrm{\hbar \vec{k}}$. The spontaneous fluorescence emission is randomised in direction and does not contribute to a net force when averaged over many scatterings. For an ion near rest, the scattering rate $\mathrm{\gamma}$ is a function of the detuning $\delta/2\pi=-14$~MHz, natural linewidth $\Gamma/2\pi=19.6$ MHz, and the laser saturation $s=P/P_{\mbox{sat}}$ and given by

\begin{equation}
\gamma= \frac{\Gamma}{2} \frac{s}{1+s+\left( 2\delta/\Gamma\right)^2}\textrm{ .}
\label{counts}
\end{equation}

\begin{figure}[h!]
\centering
\includegraphics[scale=1]{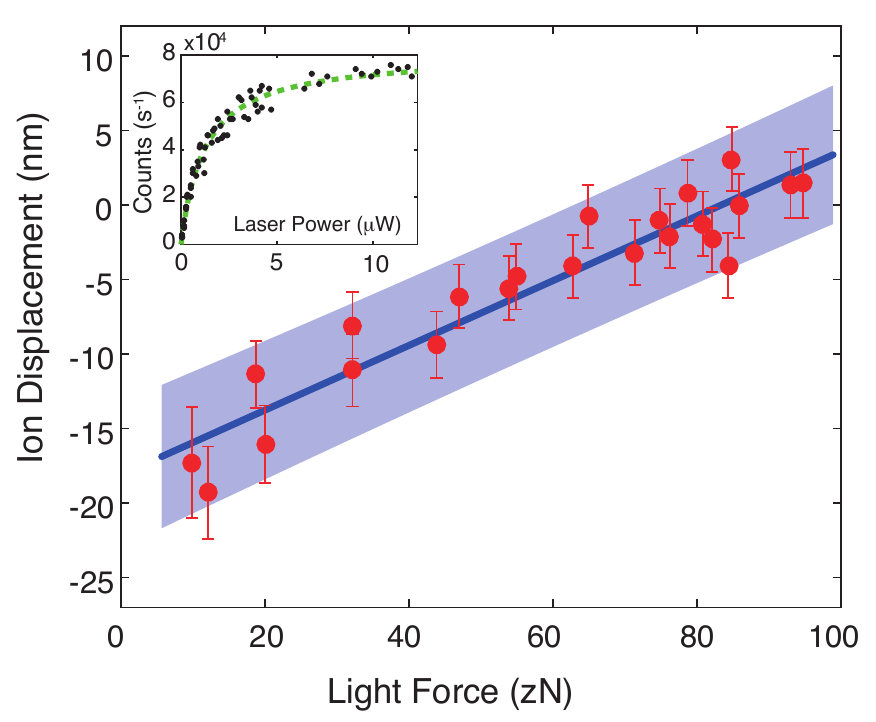}
\caption{\textbf{Light force detection with a single ion.} Ion displacement as a function of the light force (red dots) and linear fit to experimental data (blue line). The shaded region shows the 95\% confidence level for linear fit to experimental data. \textbf{Inset} Photons scattered by the ion as a function of the laser power. The green dashed line is the fitted curve from eq.~\ref{counts}, scaled by the detection efficiency, used for calculating the saturation parameter $s$ as a function of the laser power. }
\label{fig:light-force}
\end{figure}

The ion was illuminated with a laser beam along the $y$ direction and the laser power P was varied to change the scattering rate and therefore the net force. The collected ion fluorescence was split 50/50 between a photomultipler tube (PMT) and the camera. The inset in Fig.~\ref{fig:light-force} shows the saturation of the PMT counts as a function of the laser power, allowing us to fit for $P_{\mbox{sat}}$ and calibrate the scattering rate.  Figure ~\ref{fig:light-force} shows the ion displacement as a function of the applied force, together with a linear fit for forces up to 95~zN. From this measurement we calculated a drifted trap frequency in the $y$-axis of 635$\pm$25~kHz. The sensitivity of the ion as a force sensor increases by lowering the trapping potential and likewise the associated spring constant and secular frequencies. In our system the reduction in trap depth is limited by technical noise, and likely aggravated by the close proximity (3~mm) of the phase Fresnel lens's dielectric surface \cite{Harlander-ion-trap-dielectric-surface-charging}.

Fundamental quantum limits to measuring the ion position arise from the finite number of photons collected \cite{Thompson-FIONA} (see Methods). We are within a factor of (24x, 87x) in the 2D imaging plane (strong, weak) and 21x in the focal direction of the quantum limits given our trapping frequencies and imaging performance. The focusing direction is potentially more susceptible to external noise as random deviations will cause a cumulative increase in the spot width, while only the residual unbalanced sum of these deviations will impact the centroid location. Similarly the strong x axis (along the axis of the needles) trap direction will be less vulnerable to wobble in the needle positions than the two weaker (y,z) confinement axes.

Matching the trap confinement aspect ratios to the imaging system by symmetrising the confinement strength in the imaging plane and weakening the confinement along the focal direction would enhance the sensitivity. In an RF Paul Trap, application of DC biasing potentials redistributes the total spring constant provided by the RF field, allowing alteration of the trap confinement aspect ratios. Imaging performance worse than the diffraction-limit accounts for a factor of 2 in the sensitivity in the transverse directions and a factor of 8 in the focus direction. We attribute most of our error to mechanical drift since our system was not specifically engineered for long-term stability at the nm scale, although the required stability could be readily achieved with a suitably designed system.

The imaging based force sensor demonstrated here combines sub-attonewton sensitivity in all three directions with nanometer-scale spatial resolution and has important implications for a number of practical areas. In particular, for use as an in-trap probe of electric fields co-trapped particle such as a biomolecule\cite{Trypogeorgos-Cotrapping} or nanoparticle\cite{Kuhlicke-nanodiamond-ion-trap}, or investigation of surface properties \cite{Harlander-ion-trap-dielectric-surface-charging}. This type of measurements is similar to the attoNewton sensitivity 2D force sensing recently demonstrated with atomic force microscopy \cite{deLepinay-AFM-2D-Nanowire-Forces,Rossi-AFM-2D-Nanowire-Forces-other}, but with substantially higher sensitivity and potential for future improvements.  This technology can be scaled up to imaging multiple ions trapped and imaged together and used for high sensitivity and high spatial resolution study of a force field distribution.

\section*{Methods}

\subsection*{Fluorescent imaging of ions.}

Our experimental apparatus consists of a $^{174}$Yb$^+$ ion confined in a Paul trap formed near the electric field zero of a three dimensional quadrupole created by RF excitation of two tungsten needles. The ion is loaded by isotope selective photo-ionisation and laser cooled to near the Doppler limit with $\lambda$= 369.5~nm radiation on the $^{2}\mbox{S}_{1/2}$ to $^{2}\mbox{P}_{1/2}$ transition.  The resulting fluorescence at $\lambda$= 369.5~nm is collected using a 0.64 numerical aperture (NA) binary phase Fresnel lens with near wavelength resolution\cite{Jechow-11} and imaged onto an Andor iXon 897 cooled EMCCD camera with a total magnification M=395.9$\pm$0.6. The fluorescence distribution was fitted with a two a dimensional Gaussian function in order to determine the ion image's centroid position and waist widths. Centroid position with $\sim$1~nm accuracy was typically obtained for exposure times of 20~s. Re-analysis of a previous reference fluorescence image in \cite{Streed-Kielpinski-2012-Single-atom-absorption-imaging} Fig. 2d, taken with the same apparatus and an exposure time of 60~s, gave similar fitting results.

Displacements of the ion parallel to the $xy$ imaging plane are directly measured from the translation in the centroid of the ion's image.  When in focus the ion image is not sensitive to motion along $z$ to first order so to sense along this axis we deliberately image the ion slightly out of focus. A displacement of the ion parallel to the optical axis of the system $z$ leads thus leads to a change in the spot size.  We calibrated the image width to $z$ displacement in our system by translating the camera along the optical axis with the ion in a fixed position within the trap. With known magnification parameters and camera translations, we were then able to determine the ion shift along $z$ from changes in the waist of its image with an accuracy of 24~nm, 1\% of the $2.4\pm0.1\mu$m depth of focus\cite{Jechow-11}. 

\subsection*{Magnification.}
Image magnification is crucial for linking a change of the ion's image position on the CCD camera to actual ion's displacement within the trap. In order to accurately calibrate the magnification of our system we measure the separation of $2$ trapped ions\cite{Jechow-11} that were aligned parallel to the $x$ by applying a negative DC voltage to the RF needles. Due to the Coulomb repulsion, if two ions are trapped simultaneously, their separation is given by
\begin{equation}
  l = \bigg(\frac{e^{2}}{8\pi^{3}\epsilon_{0}m\nu^{2}}\bigg)^{1/3}\,=\,4.6\pm0.005\,\mu m,
\label{eq:mag_cal}
\end{equation}
where $e$ is the electron charge, $\epsilon_{0}$ is the permittivity of free space, $m$ is the mass of each $^{174}$Yb$^+$ ion and $\nu=643\pm1$~kHz is the secular frequency of the ions along the axis of separation. The distance between ions measured from the camera was 114$\pm$0.1~pixels or 1824.5$\pm$1.7~$\mu$m in the image plane. From these data we calculated a magnification of M=395.9$\pm$0.6. The magnification obtained from this method was validated by mechanically translating the RF needles $1\mu$m and verifying the ion image was displaced by the expected amount.

\subsection*{DC electric field application.}
Inside the vacuum chamber there are three additional electropolished stainless steel electrodes 2~mm away from the ion for balancing out residual DC electric fields. We applied an additional voltage on one of these electrodes to impose a small external DC field, which did not measurably shift the trapping frequencies. The ion's displacement was measured by taking images with and without the applied voltage in order to measure the effect of the external force while also capturing the background drift of the ion position. Care was taken to ensure high stability and low drift in the applied voltage source.

\subsection*{Spring constants.}
We determine the spring constants $k_i=m\omega_i^2$ of our trap by measuring the secular confinement frequencies $\omega_i$ in the three principle axes of our trap through a resonance heating method \cite{Norton-11}. The mass $m$ of $^{174}$Yb$^+$ is presumed to be constant and has been measured \cite{Rana:2012hg} with a precision of $<2\times10^{-10}$. The measured trap frequencies were at 800$\pm$1~kHz, 829$\pm$1~kHz, and 1601$\pm$1~kHz, corresponding to spring constant values of 7.29$\pm$0.02~zN/nm, 7.83$\pm$0.02~zN/nm, and 29.22$\pm$0.04~zN/nm respectively. The largest value of the spring constant corresponds to the direction parallel to the trapping needles $x$, where the confinement is stronger and the trapping frequency is higher. Our 3D quadrupole trap is nearly cylindrically symmetric, and lacking sufficient additional electrodes to break the symmetry in a known direction we were unable to determine the orientation of the two weaker trapping axes. This introduces a constrained systematic uncertainty of $\pm$7\% along those two axes. While we measured the trap frequencies to $\pm$1~kHz, measurements taken on different days showed a drift of the frequencies of the order of $\pm$15\%. This is likely due to variations in the RF electrode needle spacing as they are coupled to nanoposistioning stages outside the vacuum system through external mechanical bellows\cite{Streed-11}. For this reason trap frequencies were collected immediately after the force measurements. Active stabilisation of the trap frequencies \cite{Johnson-Ion-Trap-Freq-Active-Stab} reduces long-term (hrs) variations to below the $10^{-5}$ level. A more mechanically ridge design combined with further proposed improvements in active stabilisation could reduce trap frequency variations to below $3\times10^{-7}$, with a realistic technical lower bound tied back to the mass measurement limitation of $2\times10^{-10}$.

\subsection*{Data acquisition routine.}
We developed a special routine of data acquisition to measure absolute change of the ion position under the influence of external forces. The drifts of ion position within the trap are of the order of tens nm an hour. This is a main consequence of the fact that, the trap used for this experiment was not designed for nm level stability. To suppress the effects of drift on the force measurements we chopped the applied force on and off, at a rate equal to $2$ times the camera integration time of 20~s, and subtracted the changing drift from the force measurements. The drift value at the data point time was calculated by linear interpolation between the two neighbouring drift points and used to determine the differential displacement of the ion position, which reflects the actual influence of applied force. Several different interpolation methods were investigated (moving spline, low order polynomial, averaging two drift points), however, they all output approximately the same results. 
\subsection*{Error budget.}
The main cause of error in our apparatus was the presence of drift, which limits the accuracy of the ion's displacement measurements. Statistical uncertainty of the Gaussian fit of ion image position on the CCD camera $\sigma_{fit}$ and the statistical uncertainty of the Gaussian fit of positions of two adjacent drift points $\sigma_{DRIFT(fit)}$ are of the order of 1~nm for the $x$ and $y$ axis and 14.3~nm for the $z$ axis. The drift error is limited by the sampling rate of the drift dynamic because of our 20~s integration time and linear interpolation used to estimate the value of the drift at a specific data point. 

We estimated the drift interpolation uncertainty $\sigma_{i,DRIFT(interpolation)}$ by taking the position of a drift point along one axis $x_{i,d}$ and calculating its distance from the linear interpolation between the two adjacent points $x_{i-1,d}$ and $x_{i+1,d}$ and dividing it by 2 since the data images are acquired in between drift measurements. The drift interpolation error for each axis measurement is the average of the $\sigma_{i,DRIFT(interpolation)}$ for the individual points. The total uncertainty for the ion displacement measurement $\sigma_{ion}$, expressed as error bars on the graphs in Fig.~\ref{fig:shift_3_axes}, is given by:
$$\sigma_{ion}^2=\sigma_{fit}^2+\sigma_{DRIFT(fit)}^2+\sigma_{DRIFT(interpolation)}^2,$$ 
and the values of the error contributions for the three axis are reported in the table below.

\begin{center}
\begin{tabular}{c c c c }
\hline
    &$x$&$y$ &$z$\\
\hline
  $\sigma_{fit}$ $\mathrm{(nm)}$&$1.1$&$1.1$&$14.3$\\
	$\sigma_{DRIFT(fit)}$ $\mathrm{(nm)}$&$1.1$&$1$&$14.3$\\
	$\sigma_{DRIFT(interpolation)}$ $\mathrm{(nm)}$&$2.4$&$9.9$&$12.7$\\
\hline
\end{tabular}
\\Average errors for the data points in Fig.~\ref{fig:shift_3_axes}.
\end{center}

\subsection*{Quantum Limits}

The specific quantum limit for the accuracy of a force measurement $\delta F_i= k_i \delta x_i$ depends on the potential accuracy of our displacement measurement $\delta x_i$ given $k_i$ for our trap are the spring constants 7.29$\pm$0.02~zN/nm, 7.83$\pm$0.02~zN/nm, and 29.22$\pm$0.04~zN/nm (strong x axis). Our imaging system has a collection efficiency of $4.2\pm1.5\%$ \cite{Streed-11}, the camera a Quantum Efficiency of 35\%, and a net observed optical system transmission of 51\% after filtering, coating losses, and other sources of attenuation. The excited state lifetime of an atomic transition sets an upper bound on the rate that photons can be scattered. At saturation intensity $s=1$ with a detuning of $\delta/2\pi=-14$~MHz from resonance and a linewidth $\Gamma/2\pi=19.6$MHz we expect to detect $N=2.4\times10^6$ photons in 20 s. 

As per reference 17 we assume a Gaussian rather than an Airy or other transfer function for simplicity. For an optical system of NA=0.64 this gives an ideal Gaussian $1/e^2$ spot radius of 184 nm (FWHM of 216 nm) and a Rayleigh range $z_R$=287 nm. At the focus, in the limit of low background noise and small pixel size, for N photons in a spot of width $w_0$ the uncertainty in the centroid location is $\Delta x = w_0/\sqrt{N}$ or 0.12 nm in 20 s. With a magnification of 40.4 nm per pixel this would be "splitting the pixel" by a factor of 340, and would likely require profiling the fabrication defects of individual pixels in the camera\cite{Pertsinidis-sub-nm-super-resolution} to achieve this precision. A signal acquisition attack rate of 0.53 nm per $\sqrt{\mathrm{Hz}}$ gives a Quantum limit to the sensitivity of $15.50\pm0.03$ zN/$\sqrt{\mathrm{Hz}}$ in the strong x axis and (3.866, 4.154)$\pm$0.011 zN/$\sqrt{\mathrm{Hz}}$ in the weaker y axis. Our measured results are thus factor of 24x off this limit in the strong axis, and 87x in the weak transverse axis.

In the focusing direction the spot width $w(z)=w_0\sqrt{1+z^2/z_R^2}$ is a function of displacement from the focus z and the Rayleigh range $z_R$. The uncertainty in the width of a Gaussian peak\cite{Valentine-Gaussian-Width} is a similarly proportional to its width and the photon count $\Delta w$ is $\Delta w = w/\sqrt{2N}$. The defocusing $z_0$ with maximum fractional change in width $\partial w(z)/ \partial z= w_0/( z_R \sqrt{2})= \mbox{NA}/\sqrt{2}$ occurs at $z_0=z_R$. Considering that there are two sources of $\Delta w$ in the x and y direction, this average reduces the uncertainty by a additional factor of $1/\sqrt{2}$. Combining together gives $\Delta z= 2 \lambda/ \left( \mbox{NA} \sqrt{N}\right)$. With $N=2.4\times10^6$ photons in a 20 s exposure we would expect to be able to split the waist measurement by $\Delta w/w$= 1/340 in 1s, $\Delta z=1.16$ nm over 20 s, and an attack rate of $\Delta z=5.2$ nm /$\sqrt{\mathrm{{Hz}}}$. Translated into force sensitivity this is $(38.0,40.8)\pm0.1$ zN/$\sqrt{\mathrm{Hz}}$ or 21x the measured sensitivity.

\vspace{+1em}

{\fontsize{7.5pt}{7.5pt}\selectfont

}

\fontsize{8pt}{8pt}\selectfont
\section*{Acknowledgements}
This work was supported by the Australian Research Council (ARC) under DP130101613 and performed as part of the CSIRO/Griffith University collaboration. E.W.S was supported by ARC Future Fellowship FT130100472. V.B. and B.N. were supported by the Australian Government Research Training Program Scholarship. M.I.H. was supported by the Griffith International RHD scholarship. S.C. was supported by a CSIRO Scholarship. M.P was supported by a CSIRO OCE Postdoctoral Fellowship. The phase Fresnel lens was fabricated by M. Ferstl at the Heinrich-Hertz-Institut of the Fraunhofer-Institut f\"{u}r Nachrichtentechnik in Germany.
  
 \section*{Author contribution statement} Experiment was conceived and designed by E.W.S., S.G., and B.G.N. Data was taken by V.B., M.I.H., M.P., B.G.N., and S.C. on an experimental apparatus constructed by E.W.S., V.B., and B.G.N. Analysis by V.B., S.G., M.P., and M.L. The manuscript was prepared by V.B., M.P., M.L., and E.W.S with input from the remaining authors. Supervision by S.G., E.W.S., and M.L.
\section*{Competing Interests} The authors declare that they have no competing financial interests.
\section*{Correspondence} Correspondence and requests for materials should be addressed to EWS ~(email: e.streed@griffith.edu.au).

\end{document}